\documentclass[english,aps,prl,twocolumn,amsmath,amssymb,showpacs,superscriptaddress,notitlepage,longbibliography]{revtex4-2}

\usepackage[T1]{fontenc}
\usepackage{color}
\definecolor{page_backgroundcolor}{rgb}{1, 1, 1}
\pagecolor{page_backgroundcolor}
\usepackage{amsmath}
\usepackage{microtype}
\usepackage{bm}
\usepackage{multirow}
\usepackage{bbding}
\usepackage{verbatim}
\usepackage{makecell}
\usepackage[normalem]{ulem}
\usepackage{enumitem}
\usepackage[marginal]{footmisc}

\makeatletter
\usepackage{amssymb}
\usepackage{graphicx}

\makeatother

\usepackage{babel}

\begin{document}
\global\long\def\figurename{Fig.}%

\title{
 Spin-Valley Locking and Pure Spin-Triplet Superconductivity in Noncollinear Antiferromagnets Proximitized to Conventional Superconductors  }

\author{Song-Bo Zhang}
\email{songbozhang@ustc.edu.cn}
\address{Hefei National Laboratory, University of Science and Technology of China, Hefei 230088, China}
\address{International Center for Quantum Design of Functional Materials (ICQD), University of Science and Technology of China, Hefei, Anhui 230026, China}

\author{Lun-Hui Hu}
\email{lunhui@zju.edu.cn}
\affiliation{Center for Correlated Matter and School of Physics, Zhejiang University, Hangzhou 310058, China}

\author{Qian Niu}
%\email{niuqian@ustc.edu.cn}
\address{Department of Physics, University of Science and Technology of China, Hefei, Anhui 230026, China}
\address{CAS Key Laboratory of Strongly-Coupled Quantum Matter Physics, University of Science and Technology of China, Hefei, Anhui 230026, China}
\address{Hefei National Laboratory, University of Science and Technology of China, Hefei 230088, China}

\author{Zhenyu Zhang}
%\email{zhangzy@ustc.edu.cn}
\address{International Center for Quantum Design of Functional Materials (ICQD), University of Science and Technology of China, Hefei, Anhui 230026, China}
\address{Hefei National Laboratory, University of Science and Technology of China, Hefei 230088, China}

\begin{abstract}  
Unconventional antiferromagnets with spin-split bands, such as noncollinear magnets and the recently discovered altermagnets, serve as new constituents to explore unconventional superconductivity. Here, we unveil a new type and previously unappreciated nature of spin-valley locking in noncollinear antiferromagnets and exploit this texture to achieve pure spin-triplet superconductivity. Using chiral antiferromagnetic kagome lattices (e.g., Mn$_3$Ge and Mn$_3$Ga) coupled to conventional $s$-wave superconductors as prototypical examples, we demonstrate that the antiferromagnetic chirality strongly favors spin-triplet pairing via superconducting proximity effect, while suppressing spin-singlet pairing in the antiferromagnets away from the interfaces. Crucially, such a long-sought spin-triplet superconducting state is established without invoking the prevailing mechanism of spin-orbit coupling or net magnetization. 
Furthermore, the spin-triplet supercurrent is resilient to both in-plane and out-of-plane Zeeman fields, which exhibits distinct superiority to Ising superconductivity, serving as a compelling experimental signature of the triplet pairing and spin-valley-locked texture. 

\end{abstract}

\maketitle

\section{Introduction}

Spin-valley locking is a fascinating phenomenon discovered in monolayer transition metal dichalcogenides (TMDs)~\cite{Mak10PRL,Splendiani10NL}, where the inherently strong spin-orbit coupling naturally locks the spin and valley degrees of freedom of an electron in momentum space~\cite{DXiao12PRL,ZhuZY11PRB,Zeng12Nnano,Mak12Nnano}. 
Such a locking effect, later termed as Ising spin-orbit coupling~\cite{ZhouT16PRB}, has been shown to favor out-of-plane spin polarization and underpin a range of exotic phenomena and applications~\cite{Xu2014Nphys,Schaibley16NRM}, including valleytronics~\cite{XD2007PRL,YaoW08PRL,DXiao12PRL,CaoT2012NC,Xu2014Nphys,Schaibley16NRM} and Ising superconductivity~\cite{JMLu15Science,Saito16Nphy,XiXX2016ising,Yuan14PRL,ZhouT16PRB}.   
More recently, it has been revealed that similar spin-resolved electronic bands can also exist in a new class of quantum materials, namely unconventional antiferromagnets, such as noncollinear magnets and altermagnets. In these systems, spin-dependent band splitting is present even without invoking spin-orbit coupling~\cite{Chen2014Anomalous,Kubler14EPL,Naka19NC,Ahn19PRB,Hayami20PRB,yuanLD20PRB,Libor20SciAdv,ma2021multifunctional,Libor22PRXLandscape,Libor22PRX2,2024arXiv240602123B}, offering fundamentally new avenues for discoveries of intriguing emergent spin dependent phenomena, including notably finite-momentum~\cite{SBZ2023arXiv} and spin-triplet superconductivity.   

Spin-triplet superconductivity refers to an unconventional type of superconductivity in which two electrons pair with a total spin of 1. Since its conceptual formulation~\cite{Anderson61PR,Berezinskiv74JETPL}, spin-triplet pairing has been widely expected to play a pivotal role in dissipationless spintronics and topological superconductivity~\cite{Read00PRB,Kitaev2001unpaired,Bergeret05RMP,Buzdin05RMP,linder2015superconducting,Eschrig2015review,Sato17IOP,LeeSH24PRL}. 
Yet, despite extensive efforts, a definitive realization of spin-triplet superconductivity remains a standing challenge. To date, only a few systems have been shown to exhibit indications of spin-triplet superconductivity, including heavy-fermion superconductors~\cite{SRan19science,jiao2020chiral,GuQQ23Nature,Aishwarya23Nature,YFLi19science} and bilayer graphene~\cite{HXZhou22Science,Lee19NC}. Hybrid structures provide a promising route to induce triplet pairing~\cite{Bergeret05RMP,Buzdin05RMP,linder2015superconducting,Keizer06Nature,Demler97prb,Eschrig2015review,Bergeret2001LongRange,Volkov03PRL,Houzet07PRB,Cottet11PRL,Eschrig08NP,Robinson10Science,Triola20review,Fyhn23PRL,HuGJ23NC,jeon2021long,jeon2023chiral,Linder09PRL,Gorkov01PRL,Bergeret13PRL,Crepin15prl,Cayao17PRB,Breunig18PRL,Cayao18prb,Fleckenstein18prb}.
In these studies, the application of a certain type of magnetization is indispensable, be it of spin or orbital origin, and often relies on the existence of effective spin-orbit coupling.

In this work, we reveal a new form of spin-valley locking—a previously unrecognized property of noncollinear antiferromagnets—and further exploit this property to achieve pure spin-triplet superconductivity. As prototypical examples, we consider chiral antiferromagnetic (cAFM) kagome lattices, such as Mn$_3$Ge and Mn$_3$Ga, proximity-coupled to conventional $s$-wave superconductors
using symmetry arguments and lattice Green’s function calculations.
We demonstrate that the antiferromagnetic chirality strongly favors spin-triplet pairing, accompanied by effective suppression of spin-singlet pairing in the cAFM components away from the interfaces. In stark contrast to prevailing approaches, this long-sought spin-triplet superconducting state emerges without invoking either spin-orbit coupling or net magnetization. 
The underlying physical origin of the triplet pairing stems from the inherent noncollinear nature of valley-locked spin texture, whose presence in a realistic system is demonstrated for the first time. 
Furthermore, we show that the triplet supercurrent remains resilient to in-plane and out-of-plane Zeeman fields, which outperforms Ising superconductivity and serves as a compelling experimental signature of the triplet state.

\begin{figure}[t]
\includegraphics[width=0.48\textwidth]{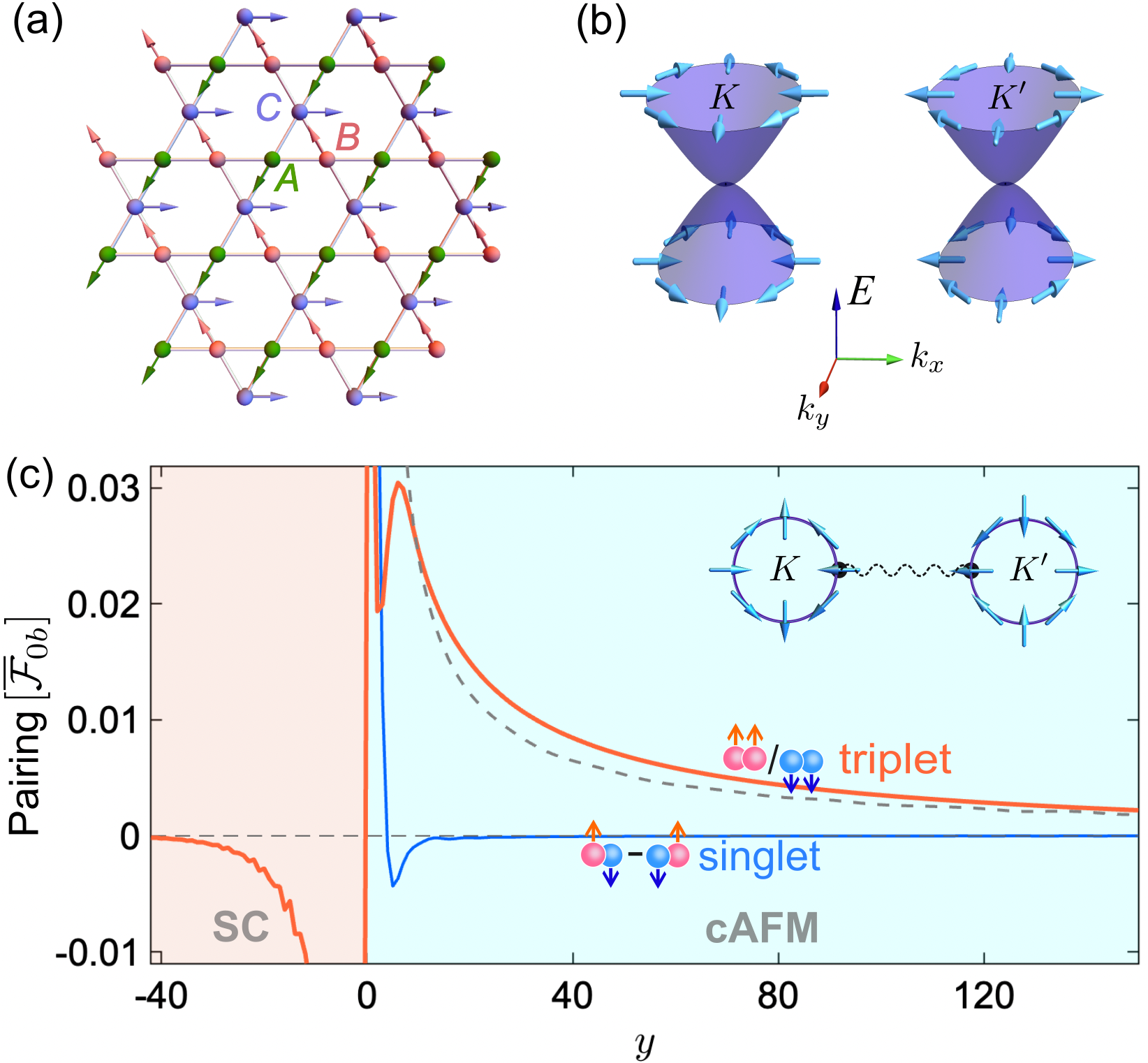}

\caption{(a) cAFM kagome lattices with local magnetic moments indicated by the arrows. 
(b) {Dirac cones at the $K/K'$ points. 
The arrows represent the in-plane spin textures, which are opposite at the $K$ and $K'$ valleys. 
(c) Spin-singlet $\overline{\mathcal{F}}_{0}$ (blue) and equal-spin-triplet $\overline{\mathcal{F}}_{t}(\equiv i\overline{\mathcal{F}}_{\downarrow\downarrow}=-i\overline{\mathcal{F}}_{\uparrow\uparrow}$, orange) pairing amplitudes in the cAFM connected to a conventional $s$-wave superconductor. The grey dashed curve represents the singlet pairing amplitude for the nonmagnetic case ($J=0$). The inset sketches  intervalley pairing that dominates at low energies. All amplitudes are averaged over the positive frequency region ($\omega\in[0,2\Delta]$) and in units of $\overline{\mathcal{F}}_{0b}$, the singlet pairing amplitude in the bulk superconductor. } 
Parameters: $\mu_S=2t$, $\mu=0.2t$, $J=0.6t$, $\Delta=0.05t$ and $k_BT=0.02\Delta$. 
}
\label{fig1:main-result}
\end{figure}

\section{Results}
\subsection{Valley-locked spin textures in cAFM}

{To elucidate the key physics, we begin with the prototypical kagome lattice model. 
In the absence of magnetism and spin-orbit coupling, the Hamiltonian is given by
\begin{eqnarray}
 \label{eq-ham0-kagome}
{\cal H}_{0}({\bf k})  = - 2t \begin{pmatrix}- 1/2& \cos({\bf k}\cdot{\bf a}_{1}) & \cos({\bf k}\cdot{\bf a}_{2})\\
\cos({\bf k}\cdot{\bf a}_{1}) & -1/2 & \cos({\bf k}\cdot{\bf a}_{3})\\
\cos({\bf k}\cdot{\bf a}_{2}) & \cos({\bf k}\cdot{\bf a}_{3}) & -1/2
\end{pmatrix}
\end{eqnarray}
in sublattice basis. Here, ${\bf k}=(k_x,k_y)$ is the in-plane momentum, $t$ denotes the hopping amplitude of $s$-electrons, and ${\bf a}_{1}=(1,0)$, ${\bf a}_{2}=(1/2,\sqrt{3}/2)$ and ${\bf a}_{3}={\bf a}_{2}-{\bf a}_{1}$ are the three nearest-neighbor vectors. The lattice constant is set to $a=1$. The energy spectrum of ${\cal H}_{0}({\bf k})$ is spin degenerate, featuring a flat band and two dispersive bands. At the band center ($E=0$), the dispersive bands touch at two inequivalent points, ${\bf K}_\pm =\pm(2/3,0)$, located at the corners of the hexagonal Brillouin zone. 
These $K/K'$ valley points obey the $D_{3h}$ little group symmetry, where the two-dimensional representation corresponds to gapless Dirac electrons with linear dispersion. 

Noncollinear magnetic order lifts the spin degeneracy~\cite{Chen2014Anomalous,Shindou01PRL}.
We consider a chiral magnetic configuration with onsite  moments ${\bf m}_{\nu}=J(\cos\theta_\nu,\sin\theta_\nu,0)
$ (with $\nu\in\{1,2,3\}$) at sublattices $\{A,B,C\}$, where $J$ is the strength, and $\theta_\nu=2(3-\nu)\pi/3$ is the orientation [Fig.~\ref{fig1:main-result}(a)], as realized in Mn$_3$Ge and Mn$_3$Ga. 
The cAFM order breaks time-reversal while preserving inversion symmetry, leading to band spin splitting~\cite{Chen2014Anomalous}.} %The vanishing net magnetization is enforced by the spin-space group~\cite{Liu2022spin}.
For small $J$ and to linear order in momentum relative to ${\bf K}_\pm$, the system can be described by the low-energy Hamiltonian:
\begin{eqnarray}
\mathcal{H}_{k\cdot p}({\bf k}) =\chi v(k_{x}\sigma_{1}+k_{y}\sigma_{2})s_{0}+J(\sigma_{1}s_{x}-\sigma_{2}s_{y})/2,
\label{eq:Kpoint}
\end{eqnarray}
where $v=\sqrt{3}t$ and $\chi=\pm 1$ labels the two valleys~\cite{SM-AFM2024}. $(s_x,s_y,s_z)$ and $s_0$ are Pauli and unit matrices in spin space, respectively, while $(\sigma_1,\sigma_2,\sigma_3)$ are Pauli matrices acting on the space spanned by two orthogonal basis states at ${\bf K}_\pm$. The fourfold degeneracy at each valley splits into two non-degenerate states and one doubly-degenerate state~\cite{Liu2022spin}.
 
Notably, the $J$-term in Eq.~\eqref{eq:Kpoint} is momentum independent, akin to Ising spin-orbit coupling term in TMDs~\cite{DXiao12PRL}. However, unlike Ising spin-orbit coupling, which preserves the out-of-plane spin component ($s_z$), this term couples to two in-plane spin components ($s_x$ and $s_y$). This distinction results in a valley-dependent coplanar spin texture [Fig.~\ref{fig1:main-result}(b)]. 
Explicitly, the energy bands of Eq.~\eqref{eq:Kpoint} are $E_{a,b}({\bf k}) = a J/2 + b \sqrt{v^2k^2+J^2/4}$ with band indices $a,b=\pm$. The corresponding spin texture can be found as
\begin{equation}
 {\bf S}_{a,b}^\chi ({\bf k}) \propto {\chi ab (\hbar/2) (\cos \theta_{\bf k}, - \sin\theta_{\bf k})},   
\end{equation}
where $\theta_{\bf k}=\arctan(k_y/k_x)$. Remarkably, each Fermi surface carries a spin winding number of $1$. The spin texture exhibits even parity, i.e., ${\bf S}_{a,b}^{+}({\bf k}) = {\bf S}_{a,b}^{-}(-{\bf k})$, due to inversion symmetry. {In real materials such as Mn$_3$Ge and Mn$_3$Ga, $J$ can be comparable to $t$~\cite{zhang2017strong,Song2024AFM}.} For small Fermi energy ($|\mu|\lesssim |J|$), each valley hosts a single Fermi surface, with \textit{opposite} spin textures, i.e., ${\bf S}_{a,b}^+ ({\bf k})=-{\bf S}_{a,b}^- ({\bf k})$ [Fig.~\ref{fig1:main-result}(b)].
The presence of the flat bands (which lie away from the band center) leads to a small energy shift ($\simeq -J^2/3t$) at the touching point, introducing an asymmetry about zero energy~\cite{SM-AFM2024}. 
All these features are corroborated by numerical results on the full kagome lattice.

\subsection{Pure spin-triplet pairing} 

The spin-valley locking profoundly affects the generation and propagation of Cooper pairs in the cAFM. When each valley hosts a single Fermi surface,
electrons at opposite momenta carry the same spin polarization. Consequently, an electron with a given spin at $\bf k$ can not find its pairing partner with opposite spin at $-\bf k$ [Fig.~\ref{fig1:main-result}(c), inset], leading to a substantial depairing effect for
spin-singlet pairs.
In contrast, this spin-valley locking promotes equal-spin-triplet pairing, making the cAFM an ideal medium for generating and purifying spin-triplet Cooper pairs.  

\begin{figure}[t]

\includegraphics[width=0.48\textwidth]{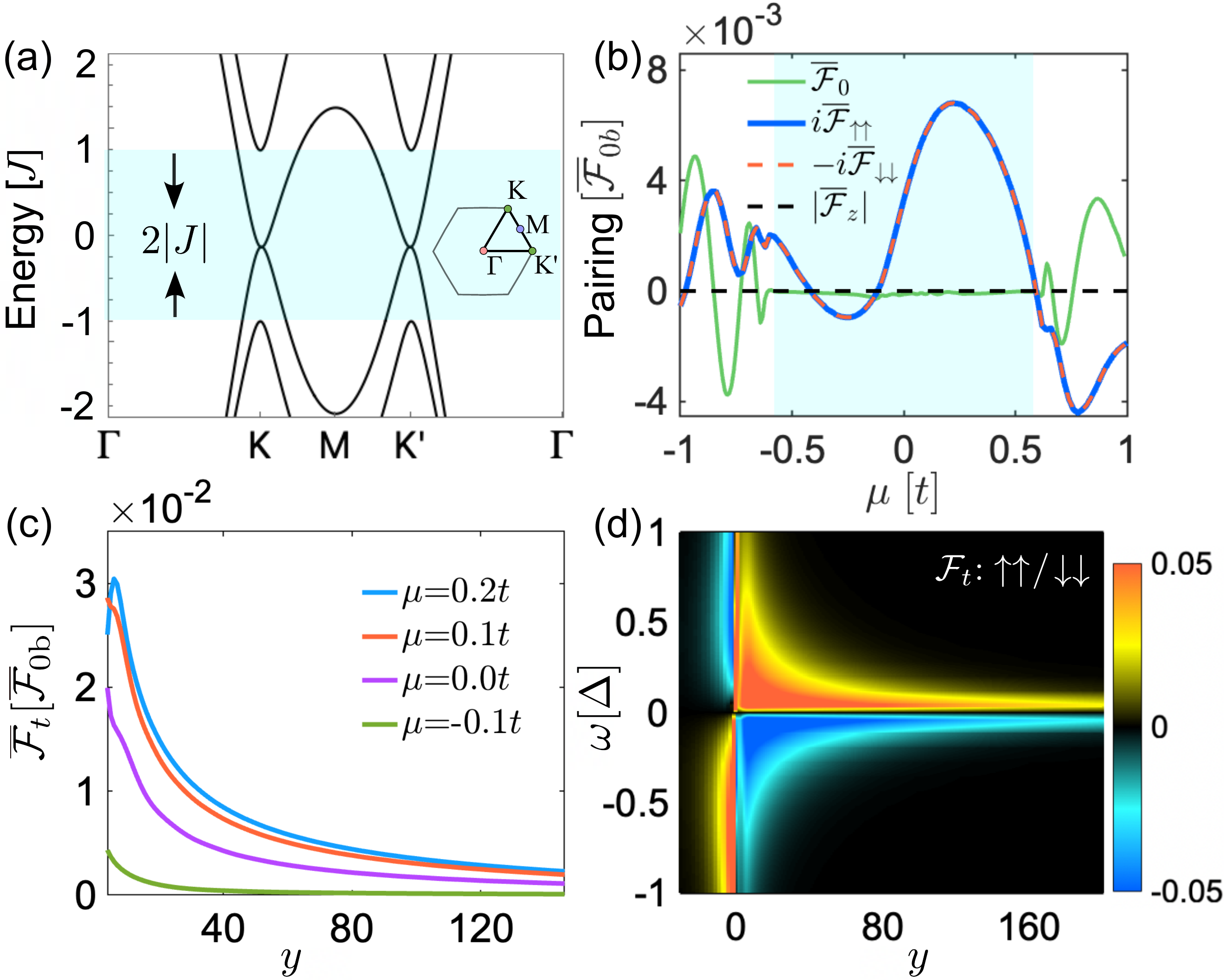}

\caption{
(a) Low-energy band structure of the cAFM lattice model with $J=0.4t$. Inset shows the Brillouin zone.
(b) Pairing amplitudes at position $y=50$ (in units of $\sqrt{3}a$) as functions of Fermi energy $\mu$.   
(c) $\overline{\mathcal{F}}_t\equiv i\overline{\mathcal{F}}_{\downarrow\downarrow}=-i\overline{\mathcal{F}}_{\uparrow\uparrow}$ as a function of $y$ for different $\mu$. (d) Phase diagram of $\mathcal{F}_t$ as a function of $\omega$ and $y$ for $\mu=0.2t$. The amplitudes are normalized by the singlet amplitude in the bulk superconductor. } 
Other parameters are the same as Fig.~\ref{fig1:main-result}(c).

\label{fig4-mu:PhaseDiagram}
\end{figure}

To demonstrate this, we consider hybrid structures where the cAFM is proximity coupled to a conventional $s$-wave superconductor along $y$-direction [Fig.~\ref{fig1:main-result}(c)]. The superconductor has an onsite pairing potential $\Delta$ and a large Fermi energy $\mu_S$ ($= 2t$) on the same lattice (but with $J=0$). The large $\mu_S$ ensures that the superconductor has quadratic bands at $\Gamma$ point, as in conventional superconductors. Assuming translation symmetry along the interface, $k_x$ remains a good quantum number. We calculate the pairing correlations from the anomalous Green's function in the Matsubara formalism using a recursive technique that iteratively constructs the Green's function by adding system layers~\cite{Sancho1984JPFMP,Sancho1985JPFMP,Asano01prb}. 
Summing over the sublattices within each unit cell and integrating over $k_x$, the pairing amplitudes are given by
\begin{align}
 \mathcal{F}_j(y,i\omega) = \int \dfrac{dk_x}{2\pi}\text{Tr} [{\bm f}_j(y,k_x,i\omega)],\;\;  
\end{align}
where $y$ represents the layer position in units of layer spacing $\sqrt{3}a$, $j\in\{0,z,\uparrow\uparrow,\downarrow\downarrow\}$ denote singlet, opposite-spin-triplet, and two equal-spin-triplet components. The singlet (triplet) pairing is even (odd) in frequency $\omega$, as required by the Pauli principle~\cite{Bergeret05RMP}. By convention, we define the spin quantization axis in $z$-direction. The phase of pairing potential is set to zero, with $\Delta = 0.05t$, and cAFM strength $J = 0.6t$ for illustration. To ensure generality, we average the pairings over the positive $\omega$ region, i.e., $\overline{\mathcal{F}}_j= (1/2\Delta) \int_0^{2\Delta} d\omega {\mathcal{F}}_j $.  
The main results are shown in Figs.~\ref{fig1:main-result}(c) and \ref{fig4-mu:PhaseDiagram}, with details of calculations in SM~\cite{SM-AFM2024}.

Due to local structural asymmetry, the junction interface induces singlet and triplet pairing. This is verified in Fig.~\ref{fig1:main-result}(c), where a mixture of singlet ($\overline{\mathcal{F}}_0$) and equal-spin-triplet ($\overline{\mathcal{F}}_{\uparrow\uparrow(\downarrow\downarrow)}$) pairings emerge.
While $\overline{\mathcal{F}}_0$ is real, the $\overline{\mathcal{F}}_{\uparrow\uparrow(\downarrow\downarrow)}$ amplitudes are imaginary and opposite in value, i.e., $\overline{\mathcal{F}}_{\uparrow\uparrow}=-\overline{\mathcal{F}}_{\downarrow\downarrow}$. In contrast, the opposite-spin-triplet pairing $\overline{\mathcal{F}}_{z}$ is vanishing. 
This implies that the triplet pairing is intrinsically spin-unpolarized, as there is no net magnetization in the system. 

Strikingly, away from the interface, inversion symmetry is restored, suppressing singlet pairing $\overline{\mathcal{F}}_0$ while allowing the triplet pairing $\overline{\mathcal{F}}_{\uparrow\uparrow(\downarrow\downarrow)}$ to penetrate the cAFM through the proximity effect. Explicitly, $\overline{\mathcal{F}}_0$ rapidly decays within an atomic-scale range and becomes negligible for $y>10$.
In contrast, the triplet pairing remains robust despite the even-parity spin splitting in the cAFM.  
It exhibits a slow decay [orange curve in Fig.~\ref{fig1:main-result}(c)], maintaining a magnitude comparable to $\overline{\mathcal{F}}_{0}$ in the nonmagnetic case [dashed curve in Fig.~\ref{fig1:main-result}(c)]. 
Thus, pure triplet pairing is maintained in the cAFM away from the interface, where it remains compatible with the cAFM order.

{We analyze the pairing amplitudes for varying Fermi energy $\mu$ in the cAFM. 
When $|\mu|\gtrsim J$, two Fermi surfaces are present at each valley. In this regime, $\overline{\mathcal{F}}_0$ and $\overline{\mathcal{F}}_{\uparrow\uparrow(\downarrow\downarrow)}$ coexist with similar magnitudes, exhibiting damped oscillations along $y$. The periodicity is determined by the splitting between the Fermi surfaces, indicative of finite-momentum Cooper pairing~\cite{SBZ2023arXiv}. 
In constrast, upon a Lifshitz transition to $|\mu|\lesssim J$, where only one Fermi surface exists per valley with a nontrivial in-pane spin texture [cyan region in Fig.~\ref{fig4-mu:PhaseDiagram}(a)], $\overline{\mathcal{F}}_0$ becomes negligible, while $\overline{\mathcal{F}}_{\uparrow\uparrow(\downarrow\downarrow)}$ remains significant. 
Figure~\ref{fig4-mu:PhaseDiagram}(b) shows the pairing amplitudes at $y=50$ (a position away from the interface) as a function of $\mu$. In both electron and hole doping regions, $\overline{\mathcal{F}}_{\uparrow\uparrow (\downarrow\downarrow)}$ remains dominant and robust,  
with its magnitude changing with $\mu$, while the decay rate remains slow  [Fig.~\ref{fig4-mu:PhaseDiagram}(c)]. The pairing amplitudes are asymmetric about $\mu=0$ due to the spectrum asymmetry of the cAFM.
A closer examination of the amplitudes as functions of frequency $\omega$ further reveals that these features persist over a broad low-energy range, with a characteristic magnitude of $\Delta$ [Fig.~\ref{fig4-mu:PhaseDiagram}(d)]. $\mathcal{F}_{\uparrow\uparrow (\downarrow\downarrow)}$ decreases to zero at high $\omega$, similar to $\mathcal{F}_0$, but its spatial decay rate appears insensitive to $\omega$. 
Additionally, we confirm the expected even (odd)-$\omega$ behavior of $\mathcal{F}_{0}$ ($\mathcal{F}_{\uparrow\uparrow(\downarrow\downarrow)}$).

We note that the non-relativistic spin-split bands and valley-locked spin texture are key gradients for realizing pure spin-triplet pairing in clean systems without relying on net magnetization, spin-orbit coupling, or intricate multilayer engineering. Such phenomena are absent in collinear antiferromagnets~\cite{Fyhn23PRL,SBZ2023arXiv,Bobkov22prb}.
In SM~\cite{SM-AFM2024}, we show that our main results also apply to lateral junctions.}

\subsection{Josephson supercurrent} 

\begin{figure}[t]

\includegraphics[width=0.48\textwidth]{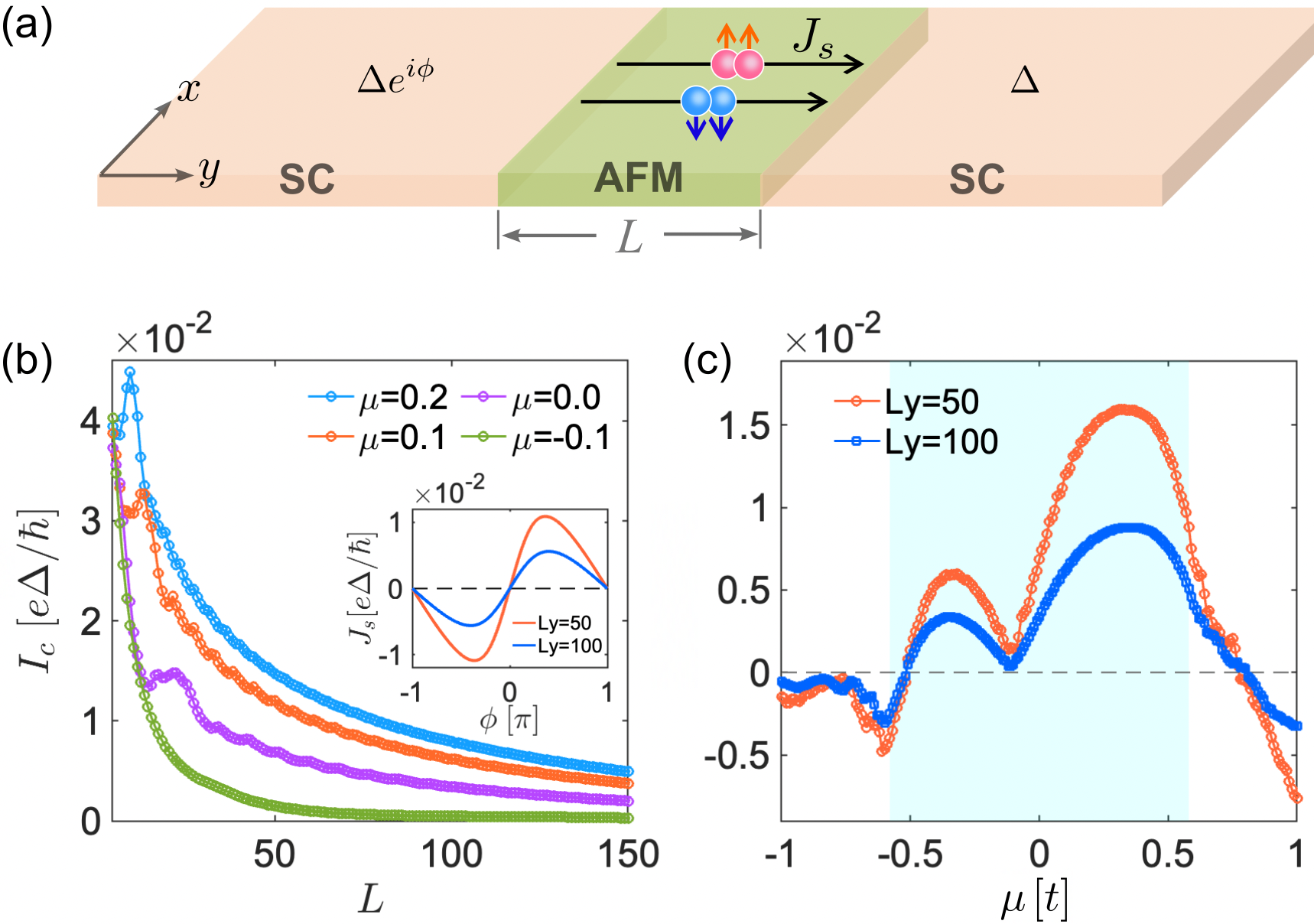}

\caption{(a) Schematic of the Josephson junction formed by two conventional superconductors sandwiching a cAFM of $L$ layers. (b) Critical supercurrent $I_c$ as a function of $L$ for different chemical potentials $\mu$ of the cAFM. Inset presents the CPRs for $\mu=0.2t$, $L=50$ and  $100$, respectively. (c) $I_c$ as a function of $\mu$ for $L=50$ and $100$, respectively. Other parameters are the same as Fig.~\ref{fig1:main-result}(c). }
\label{fig5:Ic}

\end{figure}

Now we discuss the detection of the spin-triplet Cooper pairs through their contribution to Josephson supercurrents. 
To this end, we consider a Josephson junction formed by sandwiching the cAFM of $L$ layers between two conventional superconductors along $y$-direction [Fig.~\ref{fig5:Ic}(a)]. The superconducting leads have the same pairing potential $\Delta$, but a phase difference $\phi$. 
We calculate the supercurrent using two well-established approaches: the Green function approach~\cite{Asano01prb,Sakurai17PRB,SBZhang20PRB} and free-energy approach~\cite{Tinkham2004Book}, which yield identical results.  

Figures~\ref{fig5:Ic}(b) and \ref{fig5:Ic}(c) present the main results.  
We observe substantial supercurrents with magnitudes similar to those in the nonmagnetic case. The current-phase relation (CPR) follows approximately the form $J_s= I_c\sin\phi$ [Fig.~\ref{fig5:Ic}(b), inset] due to large Fermi surface mismatch (i.e., $\mu_S=2t$ and $\mu=0.2t$), where $I_c$ denotes the critical supercurrent. 
Figure~\ref{fig5:Ic}(b) displays $I_c$ as a function of junction length $L$ for different $|\mu|(<J)$. It shows that
$I_c$ oscillates for small $L$ and then decays slowly as $L$ further increases.  
Intriguingly, the $I_c$-$L$ relation has a similar trend as $\mathcal{F}_{\uparrow\uparrow(\downarrow\downarrow)}$ with respective to $y$ [Fig.~\ref{fig4-mu:PhaseDiagram}(c)]. 
As  $\mu$ moves away from the Dirac point, both $\mathcal{F}_{\uparrow\uparrow (\downarrow\downarrow)}$ and $I_c$ [Fig.~\ref{fig5:Ic}(b)] increase simultaneously.
This correlation is further supported by Fig.~\ref{fig5:Ic}(c) [in comparison with Fig.~\ref{fig4-mu:PhaseDiagram}(b)], which shows $I_c$ as a function of $\mu$ for $L=50$ and $100$. 
These consistencies between the supercurrent and triplet pairing indicate that the supercurrent is predominantly carried by the triplet Cooper pairs.

\begin{figure}[t]

\includegraphics[width=0.48\textwidth]{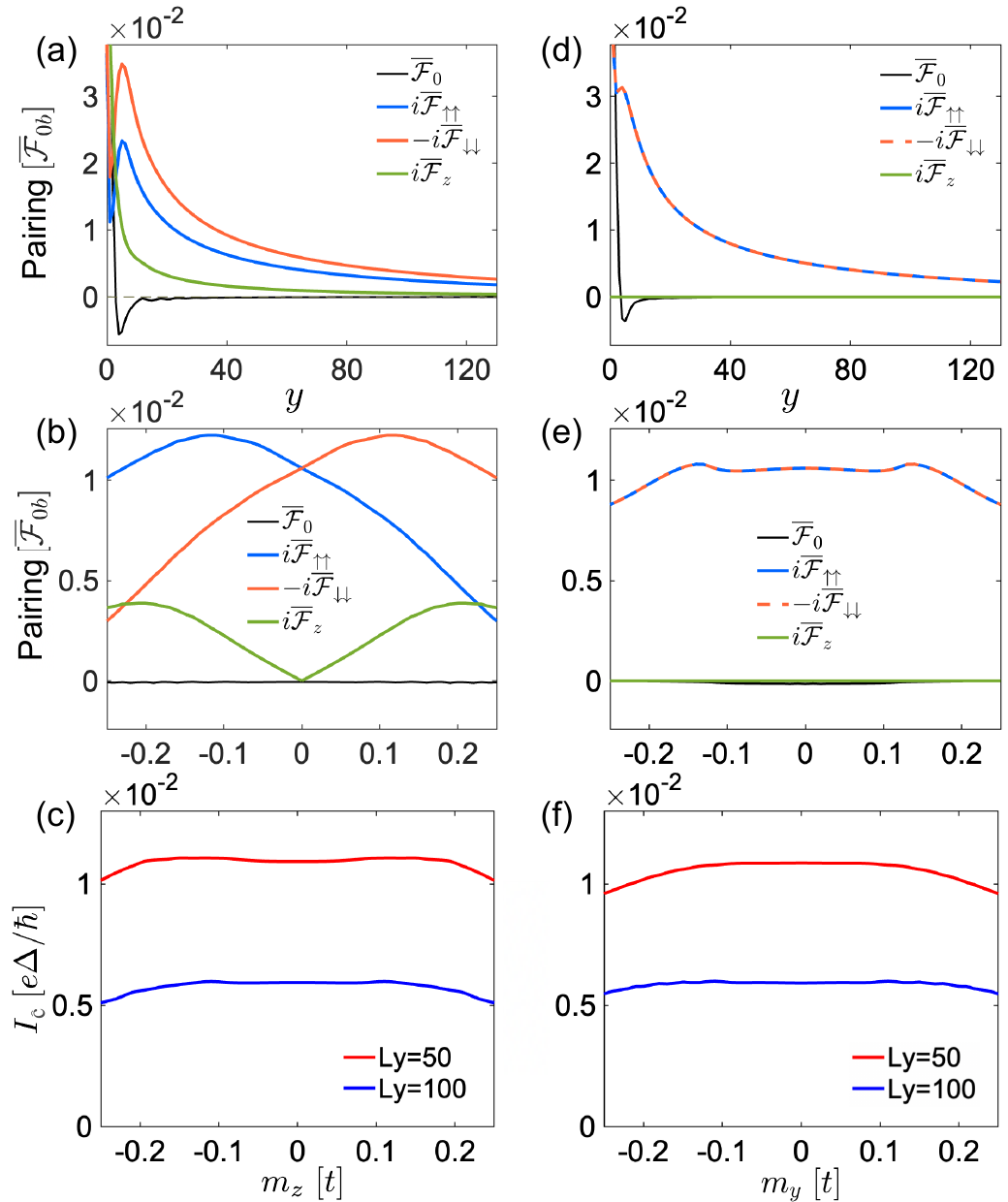}

\caption{(a) Pairing amplitudes as functions of position $y$ in the cAFM for Zeeman field $m_z=0.1t$ in $z$-direction. (b) Pairing amplitudes at $y=30$ as functions of $m_z$.
(c) Critical supercurrent $I_c$ as a function of $m_z$ for $L=50$ (red) and $L=100$ (blue). (d)-(f) Same as (a)-(c) but for Zeeman field $m_y$ in $y$-direction. Other parameters are the same as Fig.~\ref{fig1:main-result}(c).  
}
\label{fig3:canting}
\end{figure}

\subsection{Magnetic resilience of supercurrent}

The supercurrent from the triplet pairs exhibits strong magnetic resilience. We first consider out-of-plane Zeeman field, modeled by an additional $m_zs_z$ term in the cAFM. 
This term may arise intrinsically from spin canting or be induced by external magnetic fields.
Moreover, orbital magnetization in the cAFM can be significant, leading to large $m_z$ under a magnetic field~\cite{HChen20PRB}.

Figure~\ref{fig3:canting}(a) shows the paring amplitudes for $m_z=0.1t$. 
The equal-spin-triplet $\overline{\mathcal{F}}_{\uparrow\uparrow}$ (red) become dominant over $\overline{\mathcal{F}}_{\downarrow\downarrow}$ (blue) for all $y$, while $\overline{\mathcal{F}}_0$ (black) remains negligible for $y>10$. Reversing $m_z$ flips the dominance. The imbalance between $\overline{\mathcal{F}}_{\uparrow\uparrow}$ and $\overline{\mathcal{F}}_{\downarrow\downarrow}$ indicates a spin-polarized supercurrent. Nevertheless, both equal-spin-triplet components decay similarly. Meanwhile, an imaginary $\overline{\mathcal{F}}_z$ also emerges, showing a similar decay behaviour. 
Figure~\ref{fig3:canting}(b) further reveals that as $|m_z|$ ($<0.12t$) increases, ${\cal F}_{\uparrow\uparrow}$ becomes stronger while ${{\cal F}_{\downarrow\downarrow}}$ weakens, following the relations $\overline{\mathcal{F}}_{\uparrow\uparrow}(m_z)=-\overline{\mathcal{F}}_{\downarrow\downarrow}(-m_z)$ and $\overline{\mathcal{F}}_z(m_z)=\overline{\mathcal{F}}_z(-m_z)$. 
Despite these changes, $I_c$
remains nearly constant [Fig.~\ref{fig3:canting}(c)]. 
For ultra-strong $|m_z|(>0.12t)$, all pairings are suppressed due to the gap opening in the band structure. Thus, $I_c$ is suppressed overall.
We also note that the critical value of $m_z/\Delta_0$ noticeably increases, ensuring an uninterrupted supercurrent even with reduced $\Delta_0$~\cite{SM-AFM2024}.  
This strong magnetic resilience supports the robust nature of the spin-triplet pairing. 

Similar magnetic resilience occurs with in-plane fields. For illustration, Figs.~\ref{fig3:canting}(d) and \ref{fig3:canting}(e) present the pairing amplitudes under an in-plane field $m_y$ in $y$-direction (modeled by a $m_ys_y$ term in the cAFM). Notably, all pairing amplitudes are even functions of $m_y$ and remain largely unchanged for $|m_y|\sim 0.2t \gg \Delta$. Unlike the out-of-plane case, $\overline{\mathcal{F}}_z$ remains zero due to the absence of net magnetization in $z$-direction. Nevertheless, $\overline{\mathcal{F}}_0$ is consistently suppressed, leaving the junction dominated by equal-spin-triplet pairing ${\cal F}_{\uparrow\uparrow(\downarrow\downarrow)}$. Consequently, the supercurrent demonstrates strong resilience to in-plane fields as well, as confirmed in Fig.~\ref{fig3:canting}(f).      

We remark that the magnetic resilience in our cAFM system differs from that of Ising superconductivity in TMDs in two key aspects. First, while Ising superconductivity is robust only against in-plane fields, our system exhibits resilience to both in-plane and out-of-plane fields. Second, the magnetic resilience in our system withstands Zeeman fields with an energy scale comparable to hopping amplitude, which can be significantly stronger than the magnetic resilience in Ising superconductivity, where the limit is determined by the pairing potential and spin-orbit coupling~\cite{JMLu15Science,XiXX2016ising}. This distinction arises from the fundamental difference between the spin-splitting induced by spin-orbit coupling (typically on the order of 10 meV) and the interaction-induced splitting from antiferromagnetic order (typically on the order of 100 meV).

\section{Conclusion and discussion}
In this work, we introduce a new type of spin-valley locking in noncollinear antiferromagnets and exploit it to realize pure spin-triplet pairing via proximity effect, without relying on spin-orbit coupling, net magnetization, or multilayer structures. We show that these distinctive phenomena can be experimentally probed through the remarkable resilience of Josephson supercurrents to both in-plane and out-of-plane Zeeman fields.  

While our above discussion focuses on electrons with $s$-orbital and the band-center regime with $|\mu|<|J| < |t|$, the coplanar spin-valley locking and resulting pure triplet pairing can also emerge in other parameter regimes (e.g., $|\mu|>|J| > |t|$, and for other orbitals such as $p$-orbitals, as we show in SM~\cite{SM-AFM2024}. This underscores the validity and generality of our results, making them broadly applicable to realistic materials. 

Our theory can be implemented to various cAFM materials that have been experimentally confirmed, including kagome compounds~\cite{manchon2019RMP,jiang2023enumeration,ren2023enumeration,xiao2023spin} such as Mn$_3$Sn~\cite{nakatsuji2015large}, Mn$_3$Ir~\cite{zhang2016giant},
Mn$_3$Pt~\cite{liu2018electrical}, Mn$_3$Ge~\cite{Kiyohara2016Giant,nayak2016large} and Mn$_3$Ga~\cite{liu2017SciRep,Song2024AFM}. Notably, Mn$_3$Ge, Mn$_3$Ga and Mn$_3$Sn~\cite{zhang2016giant} each host one Fermi surface per valley over broad energy ranges around $E=-0.15$ eV, $0.1$ eV and $-0.1$ eV relative to the Fermi level, respectively~\cite{zhang2016giant,Qin22AM}. Moreover, Josephson junctions based on ungated Mn$_3$Ge thin films have recently become experimentally accessible~\cite{jeon2021long,jeon2023chiral}, although the regime of spin-valley locking and the effects of Zeeman fields remain unexplored. With appropriate electric gating, these materials present promising platforms to test our predictions.\\

\begin{acknowledgments}
We thank Xi Dai, Jinxing Hou, Xiaoxiong Liu, Shun-Qing Shen, Ziqiang Wang, Bin Xiang and Jinsong Xu for valuable discussions.
This work is supported by the Innovation Program for Quantum Science and Technology (Grant No.~2021ZD0302800). S.B.Z. is supported by the start-up fund at HFNL. L.H.H. is supported by the start-up of Zhejiang University and the Fundamental Research Funds for the Central Universities (Grant No. 226-2024-00068)
\end{acknowledgments}

\bibliography{ref}

\end{document}